\documentclass[journal]{IEEEtran}
\usepackage[english]{babel}
\usepackage{amsmath, amssymb}
\usepackage{graphicx,psfrag,epsf}
\usepackage{enumerate}
\usepackage[sort, numbers]{natbib}
\usepackage{url} 
\usepackage{here}
\usepackage{setspace}
\usepackage{multirow}
\usepackage{lmodern}
\usepackage{bm}
\usepackage{pdflscape}
\usepackage{authblk}
\usepackage{color}
\usepackage{subfigure}
\usepackage{booktabs,caption}
\usepackage{afterpage}
\usepackage[T1]{fontenc}
\usepackage{textcomp}
\usepackage{float}
\usepackage{enumerate}
\usepackage{rotating}
\usepackage{comment}
\usepackage{listings}

\newcommand{\btheta}{ \mbox{\boldmath $ \theta $} }
\usepackage[flushleft]{threeparttable}
\ifCLASSINFOpdf
\else
\fi
\begin{document}
%
\title{Conjugate Nearest Neighbor Gaussian Process Models for Efficient Statistical Interpolation of Large Spatial Data}
%
%
%

\author{Shinichiro~Shirota,
        Andrew~O.~Finley,
        Bruce~D.~Cook,
        and~Sudipto~Banerjee
\thanks{S. Shirota is with the RIKEN Center for Advanced Intelligence Project, Tokyo, 103-0027, Japan. S. Shirota e-mail: shinichiro.shirota@gmail.com.}
\thanks{A.O.~Finley is with the Department of Forestry, Michigan State University with Anonymous University, East Lansing, MI, 48823.}
\thanks{Bruce~D.~Cook is with the National Aeronautics and Space Administration, Goddard Space Flight Center, Greenbelt Rd, Greenbelt, MD 20771.}
\thanks{S.~Banerjee is with the Department of Biostatistics, University of California, Los Angeles, CA, 90095 USA.}
\thanks{Manuscript received XXXX; revised XXXX.}}

\maketitle
\begin{abstract}
A key challenge in spatial statistics is the analysis for massive spatially-referenced data sets. Such analyses often proceed from Gaussian process specifications that can produce rich and robust inference, but involve dense covariance matrices that lack computationally exploitable structures. The matrix computations required for fitting such models involve floating point operations in cubic order of the number of spatial locations and dynamic memory storage in quadratic order. Recent developments in spatial statistics offer a variety of massively scalable approaches. Bayesian inference and hierarchical models, in particular, have gained popularity due to their richness and flexibility in accommodating spatial processes. Our current contribution is to provide computationally efficient exact algorithms for spatial interpolation of massive data sets using scalable spatial processes. We combine low-rank Gaussian processes with efficient sparse approximations. Following recent work by \cite{Zhangetal(18)}, we model the low-rank process using a Gaussian predictive process (GPP) and the residual process as a sparsity-inducing nearest-neighbor Gaussian process (NNGP). A key contribution here is to implement these models using exact conjugate Bayesian modeling to avoid expensive iterative algorithms. Through the simulation studies, we evaluate performance of the proposed approach and the robustness of our models, especially for long range prediction. We implement our approaches for remotely sensed light detection and ranging (LiDAR) data collected over the US Forest Service Tanana Inventory Unit (TIU) in a remote portion of Interior Alaska. \end{abstract}


%
\IEEEpeerreviewmaketitle

\section{Introduction}
%
%
%
%
\IEEEPARstart{S}{tatisticians} and data scientists working on spatial data analysis in remote sensing contexts are increasingly confronting massive data sets collected over locations numbering in at least tens of millions. The primary goal often is to carry out spatial predictions at arbitrary locations while accounting for spatially varying predictors and inherent spatial dependencies. Spatial regression models incorporating spatial processes are rich and flexible, but computationally expensive and struggle to scale up to data sets with locations in the order of $10^4$---let alone $10^6$ which is typical in remote sensing data applications. There is already a substantial literature on modeling large, even massive, spatial data sets. Insightful reviews of this literature from different perspectives can be found in \cite{SunLiGenton(12)}, \cite{Banerjee(17)} and \cite{HeatonContest(18)}. 
The ``contest'' paper by \cite{HeatonContest(18)} compared the efficiency of predictive inference for a variety of scalable approaches with application in statistical \emph{gap-filling} of remotely sensed data. While some discrepancies between existing methods were noted, most of them were found to be competitive in their inferential performance. 

The work presented here was motivated by the practical need to provide computationally efficient prediction with full uncertainty quantification of light detection and ranging (LiDAR) variables for Interior Alaska as part of a National Aeronautics and Space Administration (NASA) Carbon Monitoring System (CMS) program. Current and anticipated sample-based LiDAR data collection campaigns, such as ICESat-2 \citep{abdalati2010, ICESAT2}, Global Ecosystem Dynamics Investigation LiDAR \citep{GEDI2014}, and NASA Goddard's LiDAR, Hyperspectral, and Thermal (G-LiHT) Airborne Imager \citep{cook2013}, provide only partial data coverage over domains of interest. Our current CMS work using G-LiHT, and that of future mapping/estimation initiatives with this and other sample-based LiDAR systems, require complete-coverage of LiDAR variable inputs and spatially-explicit uncertainty quantification, hence the need for statistically robust and computationally tractable prediction methods. Such methods should accommodate potentially non-stationary spatial processes which we anticipate for large regions such as the Interior Alaska study area. 

Our contribution lies in providing a practical approach for implementing classes of so called ``full scale approximation'' models that have been explored in \cite{SangHuang(12)} and, more recently, in \cite{Zhangetal(18)}. More specifically, we avoid more expensive iterative algorithms such as Markov chain Monte Carlo (MCMC), such as in \cite{Zhangetal(18)}, and formulate conjugate Bayesian models by modeling the response itself as a scalable ``sparse plus low rank Gaussian process'' (SLGP) to carry out inference (including estimation and prediction with uncertainty quantification) using exact distribution theory. This is especially beneficial for massive data sets of the magnitude we consider here. To achieve this we combine posterior inference with $K$-fold cross-validation, an approach that has been seen to be effective for massive data sets in \cite{Finleyetal(18)}. 

Following \cite{Zhangetal(18)}, we express the original spatial process as a sum of two processes: a low-rank Gaussian predictive process (GPP) \citep{Banerjeeetal(08)} and a sparsity-inducing Nearest-Neighbor Gaussian Process (NNGP) \citep{Dattaetal(16a)} for the residual process. The low-rank process captures long-range dependence and smoother variations, while the residual process captures variations at finer scales. We note that full scale approximation models have been formulated in alternative ways. \cite{Finleyetal(09a)}, \cite{SangHuang(12)}, \cite{Katzfuss(17)} and \cite{Zhangetal(18)} all use the GPP for the low-rank component but differ in how they approximate the residual process. \cite{Finleyetal(09a)} approximate the residual process using an independent process for adjusting biases in the variance, \cite{SangHuang(12)} use covariance tapering to introduce shorter-range dependence in the residual process, \cite{Katzfuss(17)} use the GPP recursively to construct a multi-resolution approximation and \cite{Zhangetal(18)} use a NNGP \citep{Dattaetal(16a)}. 

The independent process in \cite{Finleyetal(09a)} is not equipped to capture short-range dependence and, hence, less efficient than the tapered residuals in \cite{SangHuang(12)}, while the covariance tapering for residuals tends to exhibit greater shorter range dependence than the original process. Furthermore, for the very large data sets, the tapered covariance matrices will still have many nonzero elements, even with a moderate or small range for tapering function, so the computational benefits of sparsity are tempered. Another related matter is that determinant computations for covariance-tapered sparse matrices, in general, can be more complicated and less suited for Bayesian inference.

More recently, \cite{MaKang(17)} propose a full scale approach with Markov random field approximations for the residual process. Their approach renders substantial computational benefits and is effective for parameter estimation, but less suited for predictive inference since the Markov random field approximations need not yield a well-defined stochastic process over the entire domain. Also, their approach is not based on a marginal covariance decomposition, i.e., decomposition into a predictive process and a residual process, but instead includes two independent processes: one for global dependence and the other for local dependence. While this specification is flexible, two independent processes might result in over-fitting, and less accurate prediction. The NNGP can help resolve some of these issues. 

The NNGP's role as an efficient Bayesian model relies on the well-established accuracy and computational scalability of an approximation attributed to Vecchia in \cite{Vecchia(88)}, which has also been demonstrated by several authors including, more recently, by \cite{Guinness(18)}. However, a criticism for Vecchia's approximation is its dependence on the order of locations \citep{Guinness(18)}. Furthermore, this influence is exacerbated especially when the random field is too smooth because distant locations can have non-negligible impact on the value of the process at a given location under strong spatial dependence. 

The format of the paper is as follows. Section~\ref{sec:slgp} outlines our full scale approximation model in the context of spatial regression models, Section~\ref{sec:implementation} presents some details on the computations pertaining to Bayesian inference, Section~\ref{sec:sim} offers simulation studies to demonstrate the recovery of parameters and spatial surface and investigate predictive performance, and Section~\ref{sec:RDA} applies our model to the forest canopy height data in Alaska.  Finally, Section~\ref{sec:SFW} offers some discussion and future work.

\section{Sparse plus low-rank Gaussian process model}\label{sec:slgp}
Consider a set of observed locations $\mathcal{S}=\{\bm{s}_{1}, \bm{s}_{2},\ldots, \bm{s}_{n} \}$ and a response variable $y(\bm{s}_{i})$ at location $\bm{s}_{i}\in \mathcal{D}\subseteq \mathbb{R}^{d}$ in a spatial regression model,  
\begin{align}
y(\bm{s}_{i})=\bm{x}(\bm{s}_{i})^{\top}\bm{\beta}+w(\bm{s}_{i})+\epsilon(\bm{s}_{i}), \quad \text{for} \quad i=1,2,\ldots, n \label{eq:spGP}
\end{align}
where $\bm{x}(\bm{s}_{i})$ is a fixed $p\times 1$ vector of known spatially-referenced predictors, $w(\bm{s}_{i})$ is a zero mean spatial process, i.e., $\bm{w}(\mathcal{S})\sim \mathcal{N}(\bm{0}, \mathbf{C}(\mathcal{S};\btheta))$ where $\mathbf{C}(\mathcal{S};\btheta)$ is a $n\times n$ spatial covariance matrix and $\epsilon(\bm{s}_{i})\sim \mathcal{N}(0, \tau^2)$ is a white noise process capturing measurement error or micro-scale variability with $\tau^2$.  

A simple computationally efficient low rank approximation is the GPP, which projects the original process $w(\bm{s})$ onto a subspace defined by the realizations of $w(\bm{s})$ on the set of knot locations $\mathcal{S}^{*}=\{\bm{s}_{1}^{*}, \bm{s}_{2}^{*},\ldots, \bm{s}_{r}^{*} \}$. The number of locations ($r$) is smaller than ($n$), the computational cost for evaluating likelihood is $\mathcal{O}(nr^2)$ and requires $\mathcal{O}(nr)$ dynamic memory. The low rank process is smoother than the original process, so the variance of the residual in (\ref{eq:spGP}) tends to be overestimated. Full scale approximations attempt to mitigate the effects of oversmoothing by decomposing $w(\bm{s})$ as a sum of a low-rank process and a finer scale residual process, i.e., $w(\bm{s}) = w_{GPP}(\bm{s}) + w_{res}(\bm{s})$. 
We develop this further by modeling $w_{res}(\bm{s})$ using an NNGP, refer to the resulting process as the SLGP, and introduce fast algorithms for estimating conjugate SLGP models.  

\subsection{Latent SLGP models}\label{sec: latent_slgp}
Like \cite{Sangetal(11)} and \cite{SangHuang(12)}, we use the GPP \citep{Banerjeeetal(08)} as the low-rank component, but unlike covariance tapering or MRFs \citep{MaKang(17)} we follow \cite{Zhangetal(18)} and model the residual process using an NNGP \citep{Dattaetal(16a)}. We first write $w(\bm{s}) \approx w_{GPP}(\bm{s}) + w_{res,NNGP}(\bm{s})$. We look closer into each of these components. 

The GPP is derived as the conditional expectation $w_{GPP}(\bm{s}) = \mbox{E}[w(\bm{s})| \bm{w}^{*}]$, where $\mathcal{S}^{\ast}=\{\bm{s}_{1}^{\ast}, \bm{s}_{2}^{\ast}, \ldots, \bm{s}_{r}^{\ast}\}$ are a set of $r$ locations or ``knots'' with $r$ considerably smaller than $n$, and $\bm{w}^{\ast}$ is an $r\times 1$ vector with elements $w(\bm{s}_i^{\ast})$. Thus, $w_{GPP}(\bm{s})$ is the orthogonal projection of the parent process $w(\bm{s})$ on to its realizations over the set of knots. Using linearity of Gaussian processes, we write $w_{GPP}(\bm{s}) = \bm{H}(\bm{s};\btheta)\bm{w}^{\ast}$, where $\bm{H}(\bm{s};\btheta) = \bm{C}(\bm{s},\mathcal{S}^{\ast};\bm{\theta})\mathbf{C}(\mathcal{S}^{\ast};\bm{\theta})^{-1}$, $\bm{C}(\bm{s},\mathcal{S}^{\ast};\bm{\theta})$ is a $1\times r$ covariance vector and $\mathbf{C}(\mathcal{S}^{\ast};\bm{\theta})$ is the $r\times r$ covariance matrix. 
A particular advantage of the GPP is that the residual process $w_{res}(\bm{s}) = w(\bm{s}) - w_{GPP}(\bm{s})$ is tractable. The residual covariance function is
\[
\Gamma(\bm{s},\bm{s}';\bm{\theta}) = C(\bm{s},\bm{s}';\bm{\theta}) - \bm{C}(\bm{s},\mathcal{S}^{\ast};\bm{\theta})\mathbf{C}(\mathcal{S}^{\ast};\bm{\theta})^{-1}\bm{C}(\mathcal{S}^{\ast},\bm{s};\bm{\theta})\;.
\]
Our SLGP specification will be completed by approximating this exact residual process by a computationally efficient NNGP. Thus, we first project the process realizations onto realizations over a smaller set of knots, and then approximate the residual using a sparsity-inducing process.

The NNGP is a sparsity inducing GP built from Vecchia-type approximations over a reference set of locations. Let $\mathcal{R} = \{\bm{r}_1,\bm{r}_2,\ldots,\bm{r}_n\}$ be an \emph{ordered} set of locations in our domain and define neighbor sets $N(\bm{r}_i)$ to be the set of its $m$ nearest neighbors from the locations that precede it, i.e., from $ \{\bm{r}_1,\bm{r}_2,\ldots,\bm{r}_{i-1}\}$. In fact, unlike the set of ``knots'' for GPPs, the reference set can be as large as needed to construct a suitable approximation for the process. The key idea here is to approximate the $n\times n$ precision matrix 
as $\mathbf{\Gamma}(\mathcal{R};\bm{\theta})^{-1} \approx (\mathbf{I} - \mathbf{A})^{\top}\mathbf{F}^{-1}(\mathbf{I}-\mathbf{A})$, where 
$\mathbf{A}$ and $\mathbf{F}$ depend on $\btheta$,
but we suppress this dependence for simplicity in the subsequent notation.
Here, $\mathbf{F}^{-1}$ is an $n\times n$ diagonal matrix with  
\begin{align}
[\mathbf{F}^{-1}]_{ii} &= \mbox{var}\{w(\bm{r}_i)| w(N(\bm{r}_i))\} \nonumber \\
&= \Gamma(\bm{r}_i,\bm{r}_i;\bm{\theta})  \\ 
& \quad - \bm{\Gamma}(\bm{r}_i,N(\bm{r}_i);\bm{\theta})\mathbf{\Gamma}(N(\bm{r}_i);\bm{\theta})^{-1}\bm{\Gamma}(N(\bm{r}_i),\bm{r}_i;\bm{\theta}) \nonumber
\end{align}
and $\mathbf{A}$ is an $n \times n$ sparse lower-triangular such that $[\mathbf{A}]_{ij} = 0$ for all $j \geq i$ and $\{[\mathbf{A}]_{ij} : j=1,2,\ldots,i-1\}$ are kriging weights of $w(\bm{r}_i)$ based upon $w(N(\bm{r}_i))$, i.e., for $i=1,2,\ldots,n$ and $j=1,2,\ldots, i-1$
\begin{align}
[\mathbf{A}]_{ij} = \bm{\Gamma}(\bm{r}_i,N(\bm{r}_i);\bm{\theta})\mathbf{\Gamma}(N(\bm{r}_i);\bm{\theta})^{-1}
\end{align}
With this set up, the NNGP is defined recursively as $w_{res,NNGP}(\bm{r}_1) = w(\bm{r}_1)$ and $w_{res,NNGP}(\bm{r}_i) = \sum_{j=1}^{i-1} \mathbf{A}_{\btheta, ij}w_{res,NNGP}(\bm{r}_i) + \eta(\bm{r}_i)$ for $i=2,3,\ldots,n$. 
This specifies the realizations over the reference set. The definition is extended to arbitrary locations using $\bm{w}_{res,NNGP}(\bm{s}) = \sum_{j=1}^{N} \mathbf{A}_{\btheta, j}(\bm{s})w(\bm{r}_i) + \bm{\eta}(\bm{s})$ for any location $\bm{s}$ outside of $\mathcal{R}$, where $\mathbf{A}_{\btheta, j}(\bm{s}) = \bm{\Gamma}(\bm{s},N(\bm{s});\bm{\theta})\mathbf{\Gamma}(N(\bm{s});\bm{\theta})^{-1}$ where $\mbox{Pa}(\bm{s})$ is the set of $m$ nearest neighbors of $\bm{s}$ from within the locations in $\mathcal{R}$. 
Sparsity is induced because the number of nonzero elements in $\mathbf{A}$ is limited to no more than $m$ in each row.   

While $w_{GPP}(\bm{s})$ is a low-rank process that yields degenerate finite-dimensional probability laws on realizations over sets where the number of spatial locations exceed the number of knots, the $w_{res,NNGP}(\bm{s})$ is a sparse full-rank process that, by construction, will always yield non-degenerate finite-dimensional probability laws. Therefore, the process $w(\bm{s}) = w_{GPP}(\bm{s}) + w_{res,NNGP}(\bm{s})$ is also non-degenerate. We will call modeling the latent process using SLGP as the latent SLGP model. This has at least one important modeling implication: we can use $w(\bm{s})$ to model the outcomes themselves. In particular, we can devise a \emph{response} SLGP model, where rather than modeling latent spatial process, or regression coefficients, we apply SLGP to model the response itself. The response model is especially convenient for constructing conjugate Bayesian SLGPs for which we can devise fast algorithms for massive data sets. 

\subsection{Conjugate SLGP models}\label{sec: response_slgp}
We build conjugate SLGP models from the marginal, or collapsed, likelihood derived by integrating out the $\bm{w}$ from the latent SLGP. This yields
\begin{align}
 y(\bm{s}_{i})&=\bm{x}(\bm{s}_{i})^{\top}\bm{\beta}+\epsilon(\bm{s}_{i}), \quad \text{for} \quad i=1,2,\ldots,n\;  
\end{align}
and $\bm{\epsilon} \sim \mathcal{N}(\bm{0}, \mathbf{C}(\mathcal{S};\bm{\theta})+\tau^2\mathbf{I})$, where $\bm{\epsilon}$ is the $n\times 1$ vector with entries $\epsilon(\bm{s}_i)$.
We define $\alpha=\tau^2/\sigma^2$ and rewrite the marginal model as $N(\bm{y}|\mathbf{X}\bm{\beta}, \sigma^2 \mathbf{M}(\mathcal{S};\btheta))$, where $\mathbf{M}(\mathcal{S};\btheta)=\mathbf{R}(\mathcal{S};\btheta)+\alpha \mathbf{I}$ and $\mathbf{R}(\mathcal{S};\btheta)$ denotes the spatial correlation matrix. We decompose $\mathbf{M}(\mathcal{S};\btheta)$ into low rank and sparse matrices, 
\begin{align}
\begin{split}
\mathbf{M}(\mathcal{S};\btheta)&=\mathbf{J}(\mathcal{S},\mathcal{S}^{*} ;\btheta)\mathbf{R}(\mathcal{S}^{*})\mathbf{J}(\mathcal{S},\mathcal{S}^{*} ;\btheta)^{\top}+\mathbf{\Omega}(\mathcal{S};\btheta) \\
\mathbf{J}(\mathcal{S},\mathcal{S}^{*} ;\btheta)&=\mathbf{R}(\mathcal{S}, \mathcal{S}^{*};\btheta)\mathbf{R}(\mathcal{S}^{*};\btheta)^{-1}, \\ \mathbf{\Omega}(\mathcal{S};\btheta)&=\mathbf{M}(\mathcal{S};\btheta)-\mathbf{J}(\mathcal{S},\mathcal{S}^{*} ;\btheta)\mathbf{R}(\mathcal{S}^{*};\btheta)\mathbf{J}(\mathcal{S},\mathcal{S}^{*} ;\btheta)^{\top}
\end{split}
\end{align}
$\mathbf{\Omega}(\mathcal{S};\btheta)$ is a residual correlation matrix.
For the conjugate SLGP, we rewrite the above model as $N(\bm{y}|\mathbf{X}\bm{\beta}+\mathbf{J}(\mathcal{S},\mathcal{S}^{*} ;\btheta)\bm{z}^{*}, \sigma^2 \mathbf{\Omega}(\mathcal{S};\btheta))$, where $\bm{z}^{*}\sim \mathcal{N}(\bm{0}, \sigma^2\mathbf{R}(\mathcal{S}^{*};\btheta))$ is defined on $\mathcal{S}^{*}$. We need to sample $\bm{z}^{*}$ on $\mathcal{S}^{*}$ in addition to $\bm{\beta}$ and $\sigma^2$. 
$\bm{z}^{*}$ can be sampled by extending the definition of $\bm{\beta}$. 
We extend the definition of $\bm{\beta}$ to $\bm{\beta}^{*}$ such as 
\begin{align}
\begin{split}
\bm{\beta}^{*}&=\begin{pmatrix}
                     \bm{\beta} \\
                     \bm{z}^{*} \\
                   \end{pmatrix}, \quad \bm{\beta}^{*}\sim \mathcal{N}(\bm{\mu}_{\beta^{*}}, \sigma^2\mathbf{V}_{\beta^{*}}) \\
\bm{\mu}_{\beta^{*}}&=\begin{pmatrix}
                     \bm{\mu}_{\beta} \\
                     \bm{\mu}_{z^{*}} \\
                   \end{pmatrix}, \quad 
\mathbf{V}_{\beta^{*}}=\begin{pmatrix}
                         \mathbf{V}_{\beta} & \mathbf{O} \\
                         \mathbf{O} & \mathbf{R}(\mathcal{S}^{*}) 
                         \end{pmatrix}
\end{split}
\end{align}
The nearest-neighbor approximation is implemented for $\mathbf{\Omega}(\mathcal{S};\btheta)$ instead of $\mathbf{M}(\mathcal{S};\btheta)$. For fixed $\{\alpha, \phi\}$, the likelihood and prior are
\begin{align}
\mathcal{IG}(\sigma^2|a_{\sigma}, b_{\sigma})\mathcal{N}(\bm{\beta}^{*}|\bm{\mu}_{\beta^{*}}, \sigma^2\mathbf{V}_{\beta^{*}})\mathcal{N}(\bm{y}|\mathbf{X}^{*}\bm{\beta}^{*}, \sigma^2 \tilde{\mathbf{\Omega}} )  
\end{align}
where $\mathbf{X}^{*}=(\mathbf{X}, \mathbf{J}(\mathcal{S},\mathcal{S}^{*} ;\btheta))$. 
The sampling algorithm and its sampling costs are essentially the same as those of conjugate NNGP in \cite{Finleyetal(18)}. The dimension of $\bm{\beta}^{*}$ is $p+r$ where $p$ is the number of parameters and $r$ is the number of knots; $\bm{\beta}^{*}$ includes the low dimensional GP $\bm{z}^{*}$ on knots $\mathcal{S}^{*}$ with covariance matrix $\sigma^2\mathbf{R}(\mathcal{S}^{*})$. In the likelihood, the covariance matrix changed to $\sigma^2 \tilde{\mathbf{\Omega}}$ from $\sigma^2 \tilde{\mathbf{M}}$ where $\tilde{\mathbf{\Omega}}$ is the nearest-neighbor approximation of $\mathbf{\Omega}$ which is the residual correlation function. 
The posterior is obtained similar to \cite{Finleyetal(18)} and is expressed as 
\begin{align}
p(\bm{\beta}^{*},\sigma^2|\bm{y})\propto \mathcal{IG}(\sigma^2|a_{\sigma}^{*}, b_{\sigma}^{*})\times \mathcal{N}(\bm{\beta}^{*}|\mathbf{B}^{-1}\bm{b}, \sigma^2\mathbf{B}^{-1})
\end{align}
where $a_{\sigma}^{*}=a_{\sigma}+n/2$, $b_{\sigma}^{*}=b_{\sigma}+\frac{1}{2}\biggl(\bm{\mu}_{\beta^{*}}^{\top}\mathbf{V}_{\beta^{*}}^{-1}\bm{\mu}_{\beta^{*}}+\bm{y}^{\top}\tilde{\mathbf{\Omega}}^{-1}\bm{y}-\bm{b}^{\top}\mathbf{B}^{-1}\bm{b} \biggl)$, $\mathbf{B}=\mathbf{V}_{\beta^{*}}^{-1}+\mathbf{X}^{*\top}\tilde{\mathbf{\Omega}}^{-1}\mathbf{X}^{*}$ and $\bm{b}=\mathbf{V}_{\beta^{*}}^{-1}\bm{\mu}_{\beta^{*}}+\mathbf{X}^{*\top}\tilde{\mathbf{\Omega}}^{-1}\bm{y}$. 
$\bm{\mu}_{\beta^{*}}, \mathbf{V}_{\beta^{*}}, \mathbf{X}^{*}, \bm{\beta}^{*}, \tilde{\mathbf{\Omega}}$ are defined above. 
The sampling cost for conjugate SLGP model is easily obtained just by replacing $p$ in \cite{Finleyetal(18)} by $p+r$. 
In addition, we need to calculate $\mathbf{J}$ and $\mathbf{\Omega}$.  

\section{Implementation details for SLGP}\label{sec:implementation}
In the sampling of $w_{res,SLGP}$, we need to evaluate $\mathcal
{N}(\bm{w}_{res,SLGP}|\bm{0}, \mathbf{\Gamma}(\mathcal{S};\btheta))$.
From the structure of $\mathbf{A}$ it is evident that $\mathbf{I}-\mathbf{A}$ is nonsingular and $\mathbf{\Gamma} = (\mathbf{I}-\mathbf{A})^{-1}\mathbf{F}(\mathbf{I}-\mathbf{A})^{-\top}$ is. 
The nonzero elements of $\mathbf{A}$ and $\mathbf{F}$ are completely determined by the matrix $\mathbf{\Gamma}$. 
Let $\texttt{a[i,j]}$, $\texttt{f[i,j]}$ and $\texttt{Ga[i,j]}$ denote the $(i,j)$-th entries of $\mathbf{A}$, $\mathbf{F}$ and $\mathbf{\Gamma}$, respectively. Note that $\texttt{f[1,1] = Ga[1,1]}$ and the first row of $\mathbf{A}$ is $\bm{0}^{\top}$. A pseudo-code to compute the remaining elements of $\mathbf{A}$ and $\mathbf{F}$ is:
\begin{align}\label{eq: pseudocode_full_gaussian}
\begin{array}{ll}
&\texttt{for(i in 1:n-1)}\texttt{ \{ } \\ 
&\quad\texttt{a[i+1,1:i] = solve(Ga[1:i,1:i], Ga[1:i,i+1])} \\
&\quad\texttt{f[i+1,i+1] = Ga[i+1,i+1]} \\
&\hspace{60pt} \texttt{- dot(Ga[i+1,1:i],a[i+1,1:i])} \\
& \texttt{\}.}
\end{array}
\end{align}
Here  $\texttt{a[i+1,1:i]}$ is the $1\times \texttt{i}$ row vector comprising the possibly nonzero elements of the $\texttt{i+1}$-th row of $\mathbf{A}$, $\texttt{Ga[1:i,1:i]}$ is the $\texttt{i}\times \texttt{i}$ leading principal submatrix of $\mathbf{\Gamma}$, $\texttt{Ga[1:i, i]}$ is the $\texttt{i}\times 1$ row vector formed by the first $\texttt{i}$ elements in the $\texttt{i}$-th column of $\mathbf{\Gamma}$,  $\texttt{Ga[i, 1:i]}$ is the $1\times \texttt{i}$ row vector formed by the first $\texttt{i}$ elements in the $\texttt{i}$-th row of $\mathbf{\Gamma}$, $\texttt{solve(B,b)}$ computes the solution for the linear system $\texttt{Bx = b}$, and $\texttt{dot(u,v)}$ provides the inner product between vectors $\texttt{u}$ and $\texttt{v}$. 
The determinant of $\mathbf{\Gamma}$ is obtained with almost no additional cost: it is simply $\prod_{\texttt{i=1}}^{\texttt{n}}\texttt{f[i,i]}$.

The above pseudocode provides a way to obtain the Cholesky decomposition of $\mathbf{\Gamma}$. 
If $\mathbf{\Gamma} = \mathbf{L}\mathbf{D}\mathbf{L}^{\top}$ is the Cholesky decomposition, then $\mathbf{L} = (\mathbf{I}-\mathbf{A})^{-1}$. 
There is, however, no apparent gain to be had from the preceding computations since one will need to solve increasingly larger linear systems as the loop runs into higher values of $\texttt{i}$. 
Nevertheless, it immediately shows how to exploit sparsity if we set some of the elements in the lower triangular part of $\mathbf{A}$ to be zero. 
For example, suppose we set at most $m$ elements in each row of $\mathbf{A}$ to be nonzero. Let $\texttt{N[i]}$ be the set of indices $\texttt{j} < \texttt{i}$ such that $\texttt{a[i,j]} \neq 0$. We can compute the nonzero elements of $\mathbf{A}$ and the diagonal elements of $\mathbf{F}$ much more efficiently as:
\begin{align}\label{eq: pseudocode_sparse_gaussian}
\begin{array}{ll}
&\texttt{for(i in 1:n-1)}\texttt{ \{ } \\
&\qquad \texttt{Pa = N[i+1] \# neighbors of i+1}\\
&\qquad\texttt{a[i+1,Pa] = solve(Ga[Pa,Pa], Ga[(i+1),Pa])} \\
&\qquad\texttt{f[i+1,i+1] = Ga[i+1,i+1]}\\
&\hspace{60pt} \texttt{- dot(Ga[(i+1),Pa], a[i+1,Pa])}\\
& \texttt{\}.}
\end{array}
\end{align}
In (\ref{eq: pseudocode_sparse_gaussian}) we solve $n-1$ linear systems of size at most $m\times m$. This can be performed in $\mathcal{O}(nm^3)$ flops, whereas the earlier pseudocode in (\ref{eq: pseudocode_full_gaussian}) for the dense model required $\mathcal{O}(n^3)$ flops. These computations can be performed in parallel as each iteration of the loop is independent of the others. 
The density $\mathcal{N}(\bm{w}_{res,NNGP}|\bm{0}, \tilde{\mathbf{\Gamma}})$ is cheap to compute since $\tilde{\mathbf{\Gamma}}^{-1}$ is sparse and $\text{det}(\tilde{\mathbf{\Gamma}}^{-1})$ is the product of the diagonal elements of $\mathbf{F}^{-1}$. 

The factorization of $\tilde{\mathbf{\Gamma}}^{-1}$ facilitates cheap computation of quadratic forms $\bm{u}^{\top}\tilde{\mathbf{\Gamma}}^{-1}\bm{v}$ in terms $\mathbf{A}$ and $\bm{F}$. 
The algorithm to evaluate quadratic forms $\texttt{qf(u,v,A,F)}$ is provided in the following pseudocode: 
\begin{align}\label{eq: pseudocode_quadratic}
\begin{array}{ll}
&\texttt{qf(u,v,A,F) = u[1] * v[1] / F[1,1]} \\
&\texttt{for(i in 2:n}) \texttt{ \{ } \\
&\qquad\texttt{qf(u,v,A,F) = qf(u,v,A,F)} \\
&\hspace{50pt}\texttt{+ (u[i] - dot(A[i,N(i)], u[N(i)]))} \\
&\hspace{50pt}\texttt{*(v[i]-dot(A[i,N(i)],v[N(i)]))/F[i,i]} \\
& \texttt{\}.}
\end{array}
\end{align}
Observe (\ref{eq: pseudocode_quadratic}) only involves inner products of $m\times 1$ vectors. So, the entire for loop can be computed using
$\mathcal{O}(nm)$ flops as compared to $\mathcal{O}(n^2)$ flops typically required to evaluate quadratic forms involving an $n\times n$ dense matrix. Also, importantly, the determinant of $\tilde{\mathbf{\Gamma}}$ is obtained with almost no additional cost: it is simply $\prod_{\texttt{i=1}}^{\texttt{n}}\texttt{f[i,i]}$.

Hence, while $\tilde{\mathbf{\Gamma}}$ need not be sparse, the density $N(\bm{w}_{res,NNGP}|\bm{0},\tilde{\mathbf{\Gamma}})$ is cheap to compute requiring only $\mathcal{O}(n)$ flops. 
The Markov chain Monte Carlo (MCMC) implementation of the NNGP model in \cite{Dattaetal(16a)} requires updating the $n$ latent
spatial effects $\bm{w}$ sequentially, in addition to the regression and covariance parameters. While
this ensures substantial computational scalability in terms of evaluating the likelihood, the
behavior of MCMC convergence for such a high-dimensional model is difficult to study and may well prove unreliable.
\cite{Finleyetal(18)} reported that, for very large spatial datasets, sequential updating of the random effects often leads to very poor mixing in the MCMC. 
The computational gains per MCMC iteration is thus offset by a slow converging MCMC. \cite{Liuetal(94)} showed that MCMC algorithms where one or more variables are marginalized out
tend to have lower autocorrelation and improved convergence behavior. 
Here we explore SLGP models that drastically reduce the parameter dimensionality of the SLGP models by
marginalizing over the entire vector of spatial random effects. 
These models are SLGP alternatives for NNGP suggested by \cite{Finleyetal(18)}
Two different variants are developed and their relative merits and demerits are assessed both in terms of computational burden as well as model prediction and inference.\\

\subsection{Algorithm for Conjugate SLGP}\label{SLGP-alg}

\noindent {\small 
\rule{\linewidth}{1pt}
	\textbf{Algorithm: Hyper parameter tuning}} \\ 
	[-8pt]\rule{\linewidth}{1pt}\\[-12pt]
\begin{itemize}
\item[1.] Fix $\alpha$ and $\phi$, split the data into $K$ folds. 
\begin{itemize}
\item[(a)] Find the collection of neighbor sets $\mathcal{N}=\{N(i,k): i=1,2,\ldots,n; k=1,2,\ldots,K \}$
\item[b] Construct $\mathbf{\Omega}[S(-k), S(-k)]$, $\mathbf{J}[S(-k), S^{*}]$ and $\mathbf{X}^{*}=(\mathbf{X}, \mathbf{J}[S(-k), S^{*}])$ for $k=1,2,\ldots,K$
In practice, we just calculate $\mathbf{\Omega}[S, S]$, $\mathbf{J}[S, S^{*}]$ one time for each $(\alpha,\phi)$. Then extract $\mathbf{\Omega}[S(-k), S(-k)]$, $\mathbf{J}[S(-k), S^{*}]$ from them.
\end{itemize}
\item[2.] Obtain posterior means for $\bm{\beta}$ and $\sigma^2$ after removing the $k$th fold of the data:
\begin{itemize}
\item[(a)] Obtain \texttt{A(k)} and \texttt{D(k)} from $\mathbf{\Omega}[S(-k), S(-k)]$ 
\item[(b)] \texttt{F=solve($\text{V}_{\beta^{*}}$,I)}; \texttt{f=solve($\text{V}_{\beta^{*}}$, $\text{$\bm{\mu}$}_{\beta^{*}}$}) 
In practice this step is calculated one time for each $(\alpha,\phi)$
\item[(c)] Solve for $(p+r)\times (p+r)$ matrix $\mathbf{B}(k)$ and $(p+r)\times 1$ vector $\bm{b}(k)$
\begin{lstlisting}[mathescape=true, basicstyle=\ttfamily\footnotesize]
for(i in 1:p+r){
  b(k)[i] = f[i] 
    + qf(X$^*$[S(-k),i],y[S(-k)],A(k),D(k))
  for(j in 1:p+r){
    B(k)[i,j] = qf(X$^*$[S(-k),i],X$^*$[S(-k),j],
      A(k),D(k)) + F[i,j]
  }
} 
\end{lstlisting}
\item[(d)] Obtain \texttt{V(k)}, \texttt{g(k)}, \texttt{$\text{a}_{\sigma}^{*}$(k)}, \texttt{$\text{b}_{\sigma}^{*}$(k)}, by the same way in the step 2.(d) of Algorithm 5 in \cite{Finleyetal(18)}
\item[(e)] \texttt{$\text{$\hat{\beta}^{*}$}$ = g(k)}, \texttt{$\text{$\hat{\sigma}^2$}$ = $\text{b}_{\sigma}^{*}$(k)/($\text{a}_{\sigma}^{*}$(k)-1)}
\end{itemize}
\item[3.] Predicting posterior means of $\bm{y}(S[k])$:
\begin{lstlisting}[mathescape=true, basicstyle=\ttfamily\footnotesize]
for(s in S(k)) {
  N(s,k) = m-nearest neighbors of s from S(-k) 
  z = Omega(s, N(s,k))
  w = solve(Omega[N(s,k),N(s,k)],z)
  $\hat{y}(s)$ = dot(x$^*$(s), g(k)) + dot(w,
     (y[N(s,k)] - dot(X$^*$[N(s,k),],g(k))))
  v0 = dot(u, gemv(V(k),u)) + 1 + alpha 
    - dot(w,z) 
  Var($\hat{y}$(s)) = b$^*$(k)v0/(a$^*$(k)-1)
} 
\end{lstlisting}
\item[4.] Compute a scoring rule over $K$ folds 
\item[5.] Cross validation for choosing $\alpha$ and $\phi$:
\begin{itemize}
\item[(a)] Repeat steps (2) and (3) for $G$ values of $\alpha$ and $\phi$
\item[(b)] Choose $\alpha_{0}$ and $\phi_{0}$ as the value that minimize the average scoring rule
\end{itemize}
\end{itemize}
\rule{\linewidth}{1pt}
	\textbf{Algorithm: parameter estimation and prediction from conjugate SLGP model} \\ 
	[-8pt]\rule{\linewidth}{1pt}\\[-12pt]
\begin{itemize}
\item[6.] Repeat step 2 with $(\alpha_{0}, \phi_{0})$ and the full data to get $(\bm{\beta}^{*}, \sigma^2)|\bm{y}$
\item[7] Repeat step 3 with $(\alpha_{0}, \phi_{0})$ and the full data to predict at a new location $\bm{s}_{0}$ to obtain the mean and variance of $y(\bm{s}_{0})|\bm{y}$.
\end{itemize}
\rule{\linewidth}{1pt}

\subsection{Implementation and computing}\label{sec:computing}
All subsequent analyses were conducted on a Linux workstation with two 18-core Xeon(R) CPU E5-2699 v3 @ 2.30GHz processors and 512 GB of memory. The NNGP and SLGP parameter estimation and prediction algorithms were programmed in \texttt{C++} and used \texttt{openBLAS} \citep{zhang13} and Linear Algebra Package (LAPACK; \url{www.netlib.org/lapack}) for efficient matrix computations. \texttt{openBLAS} is an implementation of Basic Linear Algebra Subprograms (BLAS; \url{www.netlib.org/blas}) capable of exploiting multiple processors. Additional multiprocessor parallelization used \texttt{openMP} \citep{openmp98} to improve performance of key steps within the algorithms. In particular, substantial gains were realized by distributing the calculation of \texttt{A} and \texttt{F}, and subsequent calls to the \texttt{qf} function over the total number of available cores using the \texttt{openMP} \texttt{omp for} directive. Cross-validation and prediction algorithms were also parallelized in a straightforward manner, i.e., each cross-validation fold and prediction for a given location are independent and hence can be spread across cores.

\section{Simulation Study}\label{sec:sim}
\begin{figure*}[!htp]
\begin{center} 
	\subfigure[]{\includegraphics[trim=0cm 0cm 0cm 0cm,clip,width=8cm]{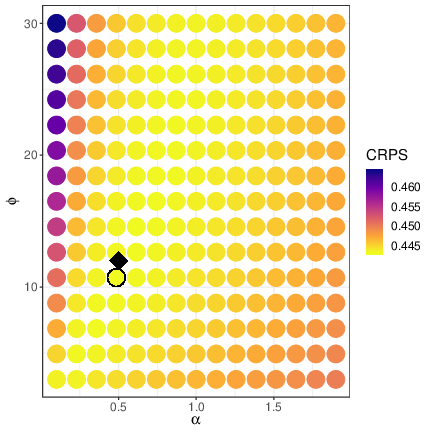}\label{sim-nngp-xval}}
	\subfigure[]{\includegraphics[trim=0cm 0cm 0cm 0cm,clip,width=8cm]{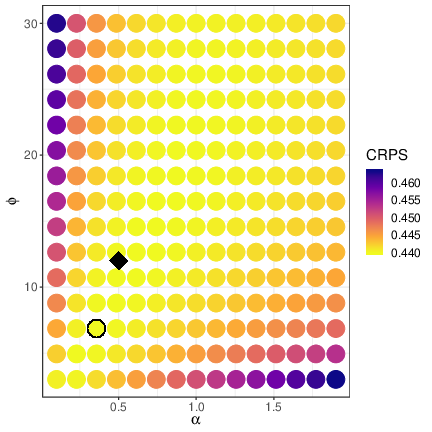}\label{sim-SLGP-xval}}
    \end{center}
    	\caption{NNGP \subref{sim-nngp-xval} and SLGP \subref{sim-SLGP-xval} CRPS search grid results for the simulated data. The ``optimal'' parameter combination, i.e., yielding the lowest CRPS, is circled. A diamond symbol identifies the ``true'' $\alpha$ and $\phi$ used to generate the data.}\label{sim-xval}
\end{figure*}
We assess the NNGP and SLGP model ability to estimate $\alpha$ and $\phi$ using a $K$-fold cross-validation and subsequent out-of-sample prediction performance following the algorithm defined in Setion~\ref{SLGP-alg}. Simulated data comprising 35,000 outcomes were generated from the Gaussian process plus nugget model within a unit square domain. The generating model was $\bm{y}\sim \mathcal{N}(\bm{0}, \mathbf{C}_{\mathcal{S}}(\bm{\theta})+\tau^2\mathbf{I})$ with an exponential covariance function $\bm{\theta} = (\sigma^2=1, \phi=12)$, and $\tau^2$=0.5, i.e., $\alpha=0.5$. These data were divided at random into a training ($n$=25,000) and holdout ($n_0$=10,000) set. Given the training set, a 5-fold cross-validation was used to find optimal values of $\phi$ and $\alpha$ under two different scoring rules; root mean squared prediction error (RMSPE) and continuous ranked probability score \cite[CRPS;][]{GneitingRaftery(07)}. While CRPS is less common for assessing predictive performance in the remote sensing literature, it has some clear advantages over the more common scoring rules such as RMSPE. As detailed in \cite{GneitingRaftery(07)}, CRPS is attractive for both practical and theoretical reasons, principal among them is that the rule favors models that yield both accurate and precise prediction, where as RMSPE only considers accuracy. Lower values of RMSPE and CRPS indicate models with better predictive performance. The search grid comprised 225 parameter sets with combinations of 15 $\alpha$ values from 0.1 to 1.9 and 15 $\phi$ values from 3 and 30. Figure~\ref{sim-xval} shows the average CRPS from the 5-fold cross-validation for each of the candidate $\alpha$ and $\phi$ pairs for NNGP and SLGP using a grid of $r$=100 knots. In Figure~\ref{sim-xval} the optimal (i.e., minimum) CRPS is circled and the ``true'' parameter value set used to generate the data is denoted with a diamond. While not shown, the RMSPE-based optimization figures looked nearly identical to Figure~\ref{sim-xval}. Figure~\ref{sim-xval} also reveals that predictive performance is fairly insensitive to choice of $\alpha$ and $\phi$ except for under fairly extreme misspecification, e.g., $\alpha \leq 0.1$ and $\alpha >1.9$ for the NNGP and SLGP and $\phi\leq3$ for SLGP (regions with purple circles). Using the 5-fold cross-validation identified optimal parameters, the NNGP and SLGP models were used to predict at the 10,000 holdout locations. Results for this out-of-sample prediction are given in Table~\ref{sim-results}, which shows negligible predictive performance between the two models regardless of the scoring rule used to select the $\phi$ and $\alpha$.

\begin{table}[!htp]
\caption{Simulated data holdout set cross-validation for RMSPE- and CRPS-based optimized parameter estimates.}
\label{sim-results}
\centering
\begin{tabular}{ccccc}
\hline
    &  \multicolumn{2}{c}{RMSPE Optim.} & \multicolumn{2}{c}{CRPS Optim.}\\
    \cmidrule(lr){2-3}    \cmidrule(lr){4-5}
    &  NNGP&  SLGP&  NNGP&  SLGP\\
 \hline
  RMSPE&0.776 & 0.775& 0.776 & 0.775\\
 CRPS&0.438 & 0.438&0.438 & 0.437\\\hline
 \hline
\end{tabular}
\end{table}

\section{Forest canopy height models}\label{sec:RDA}

Digital maps of forest structure are key inputs to many ecosystem and Earth system modeling efforts \citep{finney04, Hurtt04, stratton06, lefsky10, klein15}. These and similar applications seek inference about forest canopy height variables and predictions that can be propagated through computer models of ecosystem function to yield more robust error quantification. Given the scientific and applied interest in forest structure, there is increasing demand for wall-to-wall forest canopy height data at national and biome scales. To date, information about canopy height has been developed from sparse samples of field measurements and complete-coverage but limited spatial extent LiDAR data \citep{lefsky10, simard11, Baccini04}. Next generation LiDAR systems capable of large-scale mapping of forest canopy characteristics, such as ICESat-2 \citep{abdalati2010, ICESAT2}, Global Ecosystem Dynamics Investigation LiDAR \citep{GEDI2014}, and NASA Goddard's LiDAR, Hyperspectral, and Thermal (G-LiHT) Airborne Imager \citep{cook2013}, sample forest features using LiDAR instruments in long transects or cluster designs (see, e.g., the strips of LiDAR in Figure~\ref{tiu-bcef}). These next generation systems yield LiDAR data over the desired large spatial extents; however, the sparseness of the LiDAR sampling designs means prediction is required to deliver the desired wall-to-wall data products.

Our goal is to create high spatial resolution forest canopy height predictions, with accompanying uncertainty estimates, for the US Forest Service Tanana Inventory Unit (TIU) that covers a large portion of Interior Alaska using a sparse sample of LiDAR data from G-LiHT. We assess the proposed conjugate regression models in two settings illustrated in Figure~\ref{tiu-bcef}: 1) the small areal extent and intensively sampled Bonanza Creek Experimental Forest (BCEF); 2) the large areal extent and sparsely sampled TIU, which contains the BCEF. 

\subsection{Study sites}
\begin{figure*}[!htp]
\begin{center} 
	\subfigure[]{\includegraphics[trim=0cm 3.5cm 0.3cm 3cm,clip,width=9cm]{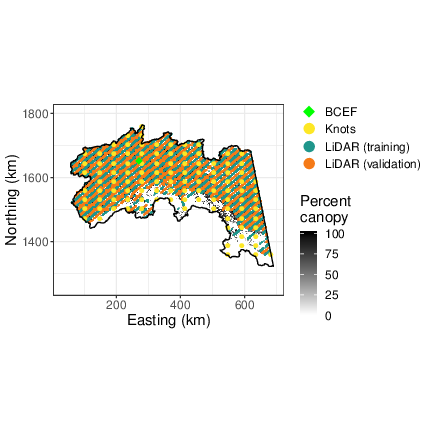}\label{tiu}}
	\subfigure[]{\includegraphics[trim=0cm 3.5cm 0.3cm 3cm,clip,width=9cm]{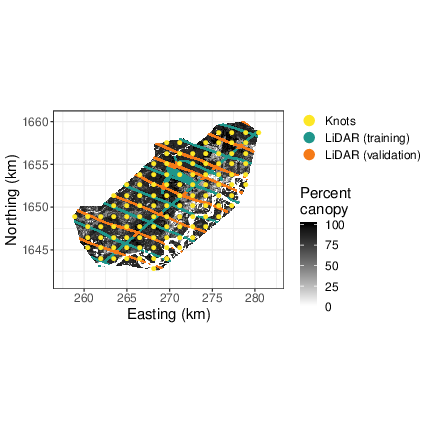}\label{bcef}}
    \end{center}
    	\caption{\subref{tiu} Tanana Inventory Unit (TIU) study area. Point symbols denote location of the Bonanza Creek Experimental Forest (BCEF) within the TIU, SLGP knots, and G-LiHT LiDAR measurements of forest canopy height for model training and validation. Underlying grayscale map is the percent tree cover across the TIU. \subref{bcef} shows location of knots, G-LiHT LiDAR forest canopy height measurements, and percent tree cover for the BCEF.}\label{tiu-bcef}
\end{figure*}

The BCEF domain delineated for this study, Figure~\ref{bcef}, is $\sim$21,000 ha and includes a section of the Tanana River floodplain along the southeastern border. Like the broader TIU, the BCEF is a mixture of non-forest and forest vegetation featuring white spruce, black spruce, tamarack, quaking aspen, and balsam poplar trees mixed with willow and alder shrubland species \cite{BCEF16}. Figure~\ref{bcef} also shows location of the $n$=188,717 G-LiHT LiDAR forest canopy height measurements.

Figure~\ref{tiu} shows the $\sim$140,000 km$^2$ TIU study area and location of the $n$=17,357,816 G-LiHT LiDAR measurements of forest canopy height. Recently, \cite{Finleyetal(18)} considered a subset of these data to assess alternate formulations of hierarchical NNGP models for improved convergence, faster computing time, and more robust and reproducible Bayesian inference. Their contribution focused on computational and parameterization improvements for NNGP models, but no effort was directed to exploring regression models beyond a basic spatially-varying intercept with an isotropic stationary covariance. The TIU's forest composition and structure are the result of myriad large and small spatial scale biotic (e.g., insect disturbance) and abiotic (e.g., soil, topography, climate, wind, fire) factors that cause spatially complex mortality and regrowth patterns. The result is a forest canopy that can exhibit both short- and long-range spatial autocorrelation and change in variability across the TIU. Such complexity in the outcome's spatial dependence surface encourages the exploration of more flexible covariance functions such as offered by the SLGP model to define the Gaussian process.

\subsection{Remote sensing data}
Forest canopy height predictions were informed using sampled LiDAR canopy height from G-LiHT (outcome variable) and complete-coverage percent tree cover and forest fire occurrence (predictor variables). G-LiHT is a portable multi-sensor system that can be mounted to a fixed wing aircraft. G-LiHT's on-board laser altimeter (VQ-480, Riegl Laser Measurement Systems, Horn, Austria) provides an effective measurement rate of up to 150 kilohertz along a 60$^{\circ}$, swath perpendicular to the flight direction using a 1,550 nanometer laser. At a nominal flying altitude of 335 m, laser pulse footprints have an approximate 10 cm diameter. The instrument is capable of producing up to eight returns per pulse. Point cloud information was summarized to a $13\times 13$ m grid cell size (grid cell area equal to 169 m$^2$). Over each grid cell, the maximum canopy height was estimated using the 100$^{\text{th}}$ percentile height of the point cloud then square root transformed before being used in the subsequent models. G-LiHT data in Summer 2014 for the study areas are available online \citep{GLIHT2016}. Two predictors that completely cover the TIU were used to help explain variability in forest canopy height. First, a Landsat derived percent tree cover data product developed by \cite{hansen13}, shown as the gray scale surfaces in Figure~\ref{tiu-bcef}. This product provides percent tree cover estimates for peak growing season in 2010 (most recent year available) and was created using a regression tree model applied to Landsat 7 ETM+ annual composites. These data are provided by the United States Geological Survey (USGS) on an approximate 30 m grid covering the entire globe \citep{hansen13}. Second, the perimeters of past fire events from 1947-2014 were obtained from the Alaska Interagency Coordination Center Alaska fire history data product \citep{AICC}. Forest recovery/regrowth following fire is very slow in Interior Alaska. Hence we discretized the fire history data to 1 if the fire occurred within the past 20 years and 0 otherwise. Fire occurrence had the same value for the entire BCEF and therefore only the percent tree cover predictor was used for the subsequent BCEF analysis.

\subsection{Analysis}
Ultimately we seek models that provide the best prediction at locations that were not sampled by the LiDAR. Using spatial regression models considered here, those locations furthest from observed locations will be most difficult to predict with high accuracy and precision. Given this consideration and the constraints of the LiDAR transect sampling design, we use the following steps to evaluate and arrive upon the ``best'' models for domain wide prediction:
\begin{enumerate}[Step 1:]
    \item Split LiDAR datasets into a training and holdout set. The holdout set consists of observations from every other LiDAR transect (to ensure a good number of distant predictions) plus a subset of observations within the transects from which the training observations are drawn (to allow for some ``nearby'' predictions), see Figures~\ref{bcef} and \ref{tiu} for BCEF and TIU respectively.
    \item Use exploratory data analysis (EDA) to identify a range of $\phi$ and $\alpha$ values over which to conduct a grid search for the ``optimal'' combination of these parameters.
    \item Search the grid defined in Step 2 to identify optimal $\phi$ and $\alpha$ for NNGP and SLGP models using 5-fold cross-validation of the training set based on CRPS. We reserve the holdout set for subsequent out-of-sampled validation in Step 4.
    \item Use NNGP and SLGP specific optimal $\phi$ and $\alpha$ from Step 3 to predict at holdout set locations, then compare these predictions to holdout set observations using CRPS. 
    \item Use optimal $\phi$ and $\alpha$ from Step 4 and all available data (i.e., training plus holdout datasets) to predict for all locations within the domain.
\end{enumerate}
Again, in practice, we might skip splitting the data into training and holdout sets and simply use $K$-fold cross-validation on all available data to identify the best model and accompanying optimal set of parameters (as illustrated in the simulated data analysis). However, given the unique LiDAR sampling design and the desire to assess how models perform making distant predictions (i.e., between LiDAR tracks), we opted for this more elaborate calibration and model choice scheme.   

\begin{figure*}[!htp]
\begin{center} 
	\subfigure[]{\includegraphics[trim=0cm 0cm 0cm 0cm,clip,width=8cm]{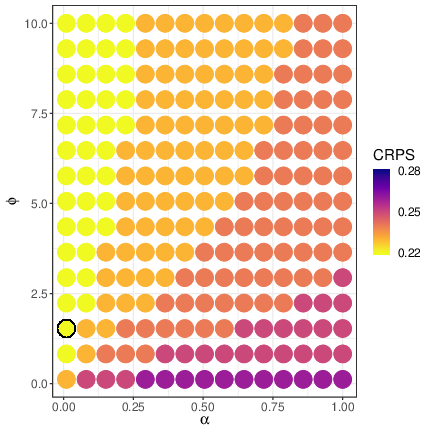}\label{becf-nngp-xval}}
	\subfigure[]{\includegraphics[trim=0cm 0cm 0cm 0cm,clip,width=8cm]{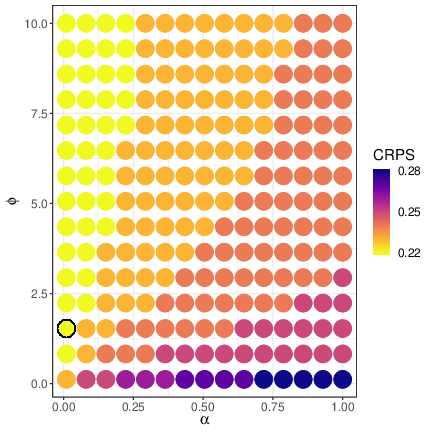}\label{bcef-SLGP-xval}}
    \end{center}
    	\caption{NNGP \subref{becf-nngp-xval} and SLGP \subref{bcef-SLGP-xval} CRPS search grid results for the BECF. The ``optimal'' parameter combination, i.e., yielding the lowest CRPS, is circled.}\label{bcef-xval}
\end{figure*}

\begin{figure}[!htp]
\begin{center} 
	\includegraphics[trim=0cm 0cm 0cm 0cm,clip,width=7cm]{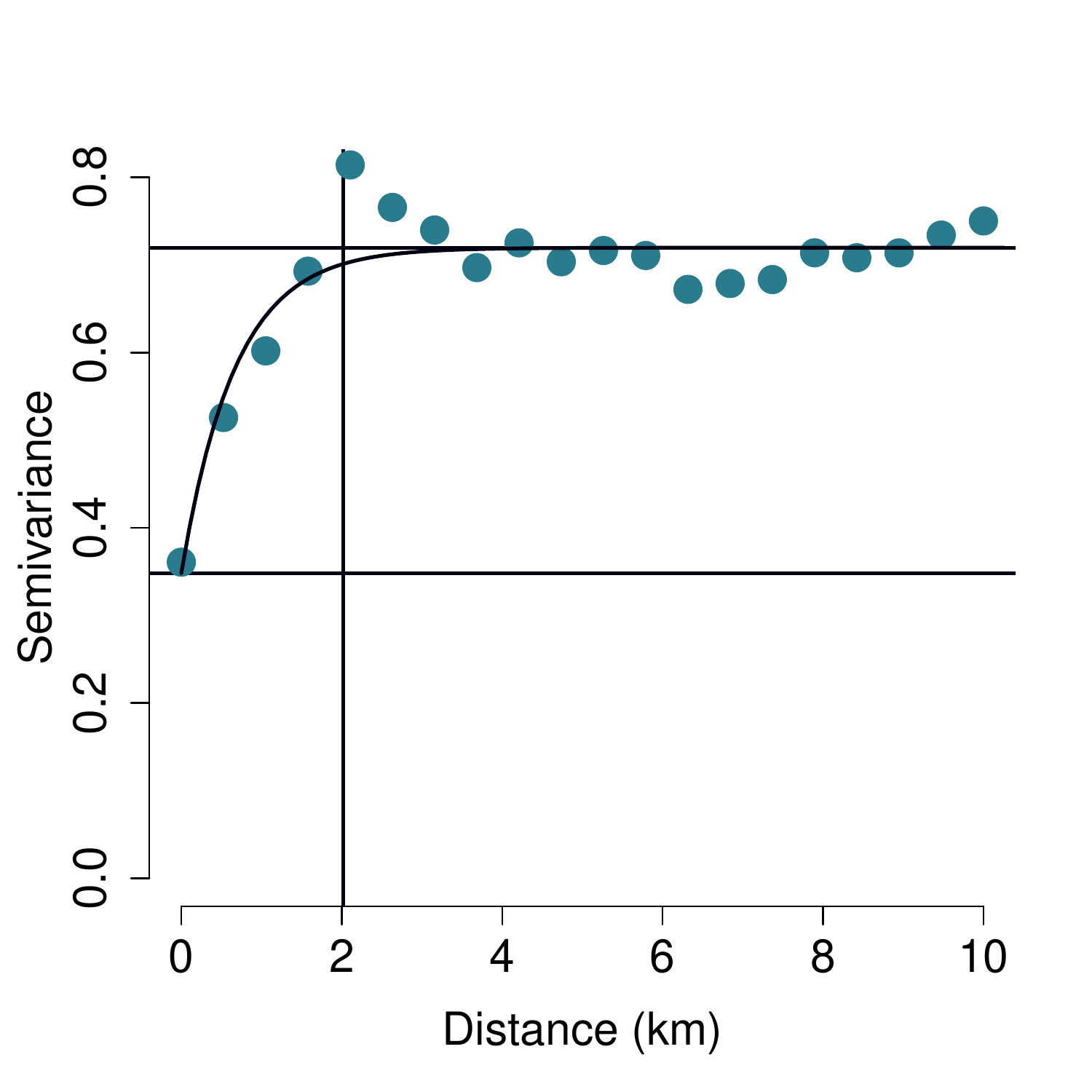}
    \end{center}
    	\caption{Semivariogram of BCEF non-spatial regression model residuals. Exponential covariance function estimate denoted by the curved line with associated estimates for $\tau^2$, $\sigma^2$, and the effective spatial range are given by the lower horizontal, upper horizontal, and vertical lines respectively.}\label{bcef-vario}
\end{figure}

For Step 2, a semivariogram of the non-spatial regression model residuals can inform how the residual spatial/non-spatial variance (i.e., outcome variance not explained by the regression mean) is partitioned and provide a rough estimate of the spatial range, see, e.g., Chapter 5 in \cite{BanerjeeCarlinGelfand(14)} for details. In the subsequent analyses we use an exponential spatial correlation function that approaches zero as the distance between locations increases. Therefore we define the distance, $d_0$, at which this correlation drops to 0.05 as the ``effective spatial range,'' which allows us to solve $\phi=-\log(0.05)/d_0$. The semivarogram and empirical parameter estimates (from a non-linear regression) for the BCEF are given in Figure~\ref{bcef-vario}. These estimates suggest a search grid over the intervals $\alpha=(0.01,1)$ and $\phi=(0.1,10)$ would be reasonable. Figure~\ref{bcef-xval} depicts this search grid and results from Step 3 that identify the ``optimal'' set $\phi$=1.53 and $\alpha=0.01$ as producing the minimum CRPS for both NNGP and SLGP. 

Following Step 4, the model specific ``optimal'' parameter set was used to predict for holdout set locations, the results of which are given in Table~\ref{bcef-holdout} along with the associated fixed and estimated model parameters. Here, in addition to the NNGP and SLGP, we included the parameters and prediction metrics for the non-spatial regression model (NS), which for the BCEF is informed using an intercept and percent tree cover predictor. Also, we considered two different knot grid intensities with which to calculate the SLGP (a depiction of the 200 knot configuration is given in Figure~\ref{bcef}). Here, CRPS and RMSPE values show the NNGP and SLGP models yield improved prediction over the NS model, and there is no appreciable difference between the NNGP and SLGP models. There are a few things to note in the parameter estimates. First, consistent with our knowledge of the BCEF's forest canopy, there is strong but localized spatial structure, which is reflected in the small noise to signal ratio $\alpha$ and short spatial range of $\sim$2 km ($-\log(0.05)/1.53$). Second, as shown by its regression parameter estimate, $\beta_{TC}$, the tree cover predictor explains some variability in canopy height. Third, compared with the empirical semivariogram estimate of $\sigma^2$, which is $\sim$0.6 in Figure~\ref{bcef-vario}, the NNGP and SLGP estimate of 3.76 seems quite large. However, this difference in spatial variance is simply due to the objective function used to estimate the parameters. The parameters in the non-spatial regression from which the residuals were derived for the semivariogram were estimated to minimize the root mean squared difference between the observations and the fitted values. In contrast the parameter estimates in Table~\ref{bcef-holdout} were effectively chosen to minimize CRPS via cross-validation. Alternately, one could choose to use minimum RMSPE as the objective function for NNGP and SLGP parameterization (the result of which is given in the last row of Table~\ref{bcef-holdout}), in which case the two models estimate $\sigma^2=0.8$ and $\tau^2=0.6$, which are much closer to the semivariogram estimates.

\begin{center}
\begin{table}[]
 \begin{threeparttable}
\caption{BCEF holdout set cross-validation CRPS-based parameter estimates and prediction metrics. Parameter variances estimates given in parentheses. RMSPE values were generated from holdout set cross-validation RMSPE-based parameter estimates that are not reported in this table.}
\label{bcef-holdout}
\addtolength{\tabcolsep}{-5pt} 
\begin{tabular}{ccccc}
\hline
    &  &    &\multicolumn{2}{c}{SLGP}\\
    &  NS & NNGP&  $r$=110  & $r$=200  \\
 \hline
 $\beta_0$  & 1.80 (3.2$^\dag$)& 2.34 (0.045) & 2.61 (0.058) & 2.71 (0.058) \\
 $\beta_{TC}$  & 0.028 (5.0$^\ast$) &  0.004 (1.0$^{\ddag}$) &  0.004 (1.0$^{\ddag}$) & 0.004 (1.0$^{\ddag}$) \\
  $\alpha$  & --& 0.01&  0.01& 0.01\\
 $\tau^2$ &0.69 & 0.04& & 0.04\\
 $\sigma^2$  &-- &  3.76 (1.1$^\dag$)& 3.76 (1.1$^\dag$)& 3.76 (1.0$^\dag$) \\
 $\phi$  & --& 1.53& 1.53& 1.53\\
  CRPS&0.49  &0.33 &0.32  &0.32 \\
 RMSPE&0.85  &0.59 &0.59  &0.59\\\hline
 \hline
\end{tabular}
\addtolength{\tabcolsep}{5pt}
   \begin{tablenotes}
      \small
      \item $^\dag$ times $10^{-3}$; $^{\ddag}$ times $10^{-5}$; $^\ast$ times $10^{-8}$. 
    \end{tablenotes}
 \end{threeparttable}
\end{table}
\end{center}

\begin{figure*}[!htp]
\begin{center} 
	\subfigure[BCEF canopy height]{\includegraphics[trim=0cm 2.5cm 0cm 2.5cm,clip,width=8cm]{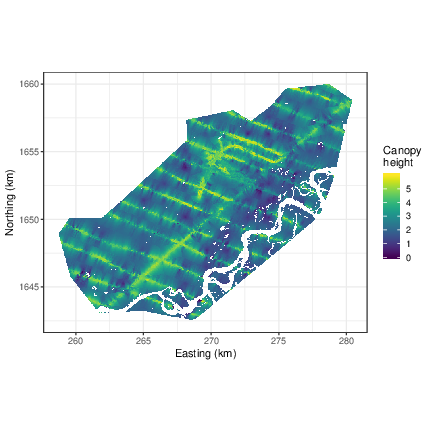}\label{bcef-yhat}}
	\subfigure[BCEF canopy height variance]{\includegraphics[trim=0cm 2.5cm 0cm 2.5cm,clip,width=8cm]{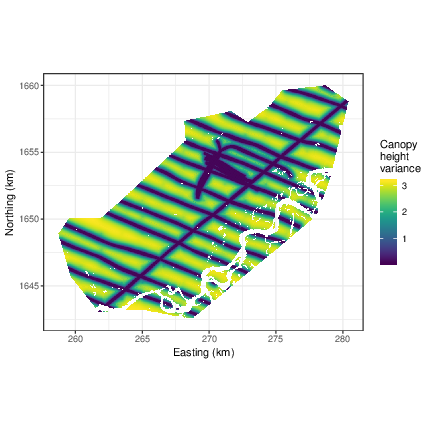}\label{bcef-yvar}}\\
	\subfigure[TIU canopy height]{\includegraphics[trim=0cm 2.5cm 0cm 2.5cm,clip,width=8cm]{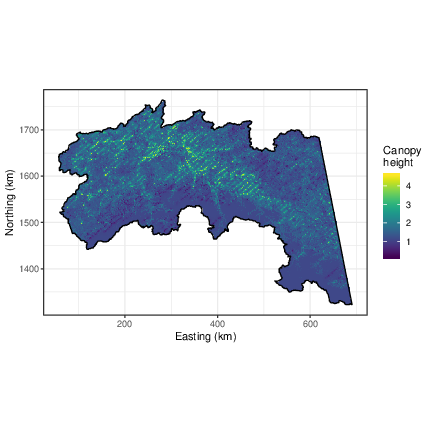}\label{tiu-yhat}}
	\subfigure[TIU canopy height variance]{\includegraphics[trim=0cm 2.5cm 0cm 2.5cm,clip,width=8cm]{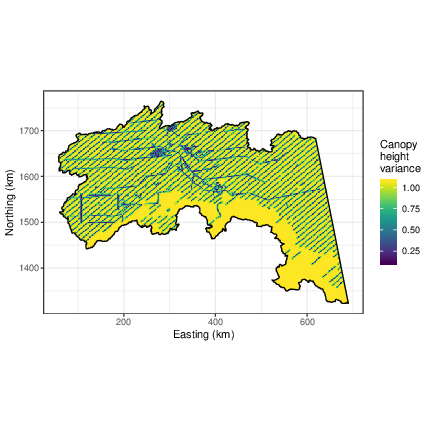}\label{tiu-yvar}}
    \end{center}
    	\caption{BCEF and TIU pixel-level SLGP model predicted square root of forest canopy height mean and associated variance.}\label{tiu-bcef-pred}
\end{figure*}

Figure~\ref{bcef-yhat} and \ref{bcef-yvar} show the predicted outcome and associated variance for every pixel across the BCEF. These prediction mean and uncertainty surfaces exhibit artifacts from the LiDAR sampling design. Notably, the mean shows more local patterns in canopy height where data is plentiful, and a retreat to the mean canopy height in regions far from observed data. The data sampling design and fairly short effective spatial range also exacerbate a feature of the NNGP and SLGP prediction algorithm. Specifically, because each prediction is informed by a set of nearest geographical neighbors, which often come from the same area along a LiDAR transect, the prediction surface has a \emph{smudged} appearance perpendicular to the transects. Compared to the NNGP, the addition of the knots in the SLGP does smooth this artifact to some degree. The uncertainty surface Figure~\ref{tiu-yvar}, clearly shows observations along and adjacent to LiDAR transects inform prediction as reflected in the substantially higher precision.

\begin{center}
\begin{table}[]
 \begin{threeparttable}
\caption{TIU holdout set cross-validation CRPS-based parameter estimates and prediction metrics. Parameter variances estimates given in parentheses. RMSPE values were generated from holdout set cross-validation RMSPE-based parameter estimates that are not reported in this table.}
\label{tiu-holdout}
\addtolength{\tabcolsep}{-5pt} 
\begin{tabular}{ccccc}
\hline
    &  &    &\multicolumn{2}{c}{SLGP}\\
    &  NS & NNGP&  $r$=196  & $r$=446  \\
 \hline
 $\beta_0$  &0.55  (1.1$^\dag$)& 1.0 (1.3$^\dag$) &  0.98 (2.0$^\dag$) &1.0 (1.2$^\dag$) \\
 $\beta_{TC}$  &0.027 (2.1$^{\ast}$) & 0.016  (1.0$^{\ast}$) &   0.017 (1.0$^{\ast}$) &0.016  (1.0$^{\ast}$) \\
  $\beta_{Fire}$  &-0.03 (9.9$^\ddag$) & 0.12  (1.0$^{\ast}$) &  0.12 (7.8$^{\ast}$) &  0.12 (1.0$^{\ast}$) \\
  $\alpha$  & --&0.13 &0.255  &0.13 \\
 $\tau^2$ & 0.69 &0.17 &0.23 &0.17 \\
 $\sigma^2$  &-- &   1.31 (7.0$^\dag$)&  0.61 (1.0$^\ddag$)&  1.31 (7.0$^\ddag$) \\
 $\phi$  & --& 0.6& 0.825 & 0.6 \\
  CRPS & 0.47 &0.38 &0.38  & 0.38 \\
 RMSPE & 0.83 &0.69 &0.69  & 0.69\\ 
\hline
\end{tabular}
\addtolength{\tabcolsep}{5pt}
   \begin{tablenotes}
      \small
      \item $^\dag$ times $10^{-3}$; $^{\ddag}$ times $10^{-5}$; $^\ast$ times $10^{-8}$. 
    \end{tablenotes}
 \end{threeparttable}
\end{table}
\end{center}

We followed the same steps for the TIU analysis. Parameter estimates and prediction performance metrics are given in Table~\ref{tiu-holdout}. Here, like the BCEF there is no notable difference between the NNGP and SLGP models in regard to their predictive performance. Both the percent tree cover and forest fire occurrence predictors explain some variability in canopy height as reflected by $\beta_{TC}$'s and $\beta_{Fire}$'s non-zero mean and small variance. The addition of the spatial random effects do improve prediction over that of the non-spatial regression. Also, we see both spatial models identify a slightly longer spatial range, i.e., $\sim$5 km but a large noise to signal variance ratio compared with the BCEF. 

Finally, we fit the NNGP and SLGP with 200 knot models using $\alpha$=0.13 and $\phi$=0.6 to the entirety of the TIU's $n$=17,357,816 observations. Using the parallelized model code and 18 CPU cores of the computer described in Section~\ref{sec:computing}, run time for the NNGP model was $\sim$9 seconds and $\sim$22 minutes for the SLGP model. Both models were used to predict canopy height for the TIU domain. Similar to the validation results presented in Table~\ref{tiu-holdout}, the resulting TIU predictions were indistinguishable. Figures~\ref{tiu-yhat} and \ref{tiu-yvar} show the SLGP model's pixel-level mean and variance predictions for the TIU. All of the same prediction features seen in the BCEF are apparent in the TIU forest canopy. 

\section{Summary and future work}\label{sec:SFW}
This work further develops Bayesian NNGP models proposed by \cite{Dattaetal(16a)} with the aim to improve approximation, predictive performance, and most importantly computational efficiency to facilitate analysis of large data sets encountered in remote sensing applications. Massive computational gains are realized by using a conjugate method  in place of MCMC-based inference for NNGP model parameter and predictive inference \cite{Finleyetal(18)}. Extending recent work by \cite{Zhangetal(18)} we propose, implement, and illustrate a conjugate sparse plus low rank approximation (SLGP) that combines a Gaussian predictive process and nearest neighbor Gaussian process model. The proposed NNGP and SLGP algorithms and software implementation takes advantage of parallel computing and thrifty memory management to deliver Bayesian kriging inference for an unprecedentedly large data set (TIU $n\approx$17 million locations) in less than a minute. The resulting forest canopy height models with associated pixel-level uncertainty quantification will help inform forest biomass and carbon models as part of an interior Alaska NASA Carbon Monitoring System. The level of computational efficiency delivered by conjugate NNGP and SLGP models opens many opportunities for previously unavailable statistically valid uncertainty assessment in remote sensing applications. 

While full posterior inference via MCMC is often preferable, it is not yet computationally feasible for massive data sets. Also, from a practical standpoint, our experience and results presented here suggest that prediction accuracy and precision often do not suffer from fixing some spatial covariance parameters at reasonable values (e.g., the spatial decay parameter $\phi$ and noise to signal variance ratio $\alpha$), hence we see value in pursuing computationally efficient NNGP and SLGP conjugate models such as those explored here. Following \cite{Zhang19}, who detail how latent process inference can also be achieved for NNGP models within a conjugate Bayesian framework, our future work will focus upon explicitly estimating latent effects for SLGP models without using MCMC. Inference on the latent process (e.g., maps) can prove useful for identifying patterns in model residual and hypothesis testing. Other work will focus on exploring conjugate NNGP and SLGP models for high-dimensional multivariate outcomes using spatial factor models \citep{Taylor-Rodriguezetal(18)}.


%



\section*{Acknowledgment}
The work of the first and third authors was supported, in part, by federal grants NSF/DMS-1513654, NSF/IIS-1562303 and NIH/NIEHS-1R01ES027027. 
The second author was supported by NSF/DMS-1513481, EF-1137309, EF-1241874, and EF-1253225.

\ifCLASSOPTIONcaptionsoff
  \newpage
\fi

\end{document}